\documentstyle[12pt]{article}
\begin{document}
\begin{titlepage}
\pagestyle{empty}
\baselineskip=21pt
\rightline{UMN-TH-1318/94}
\rightline{hep-ph/yymmddd}
\rightline{November 1994}
\vskip .2in
\begin{center}
{\large{\bf Nucleosynthesis and the
Time Dependence of \\
 Fundamental Couplings}} \end{center}
\vskip .1in
\begin{center}
Bruce A. Campbell

{\it Department of Physics, University of Alberta}

{\it Edmonton, Alberta, Canada T6G 2J1}

and

Keith A. Olive

{\it School of Physics and Astronomy, University of Minnesota}

{\it Minneapolis, MN 55455, USA}

\vskip .1in

\end{center}
\vskip .5in
\centerline{ {\bf Abstract} }
\baselineskip=18pt
We consider the effects of the time dependence of couplings due to their
dependence on a dilaton field, as occurs in superstring theory, as well as
in gravity theories of the Jordan-Brans-Dicke type. Because the scale
parameters of couplings set by dimensional transmutation depend
exponentially on the dilaton vev, we may obtain stringent limits on the
shift of the dilaton from the requirement that the induced
shift in the couplings not vitiate the successful calculations of
element abundances for big-bang nucleosynthesis. These limits can
be substantially stronger than those obtained directly from the
dilaton-induced change in the gravitational coupling.

\noindent
\end{titlepage}
%\newpage
\baselineskip=18pt

\def\la{{{\lower 5pt\hbox{$<$}} \atop {\raise 5pt\hbox{$\sim$}}}~}
\def\ga{{{\lower 2pt\hbox{$>$}} \atop {\raise 1pt\hbox{$\sim$}}}~}
\def\mtw#1{m_{\tilde #1}}
\def\tw#1{${\tilde #1}$}
\def\beq{\begin{equation}}
\def\eeq{\end{equation}}

The successful predictions of the light element abundances in the standard
model of big bang nucleosynthesis (SBBN) \cite{wssok} provides a basis to
test extensions to the standard model of particle interactions.
While deviations to SBBN typically induce changes in all of the light
element abundance predictions (D, $^3$He, $^4$He, and $^7$Li), particle
physics models are mostly constrained by the $^4$He mass fraction, $Y_p$.
In the SBBN, the abundances are primarily sensitive to only a single
parameter, the baryon-to-photon ratio, $\eta$. Consistency between
 the predictions of SBBN and the observational determinations of the
light element abundances restricts $\eta$ to a narrow range between
$2.8 \times 10^{-10}$ and $4 \times 10^{-10}$. In this range the calculated
$^4$He  mass fraction lies in the range $Y_p = 0.239 - 0.246$ \cite{kern}.
This is slightly high when compared with the observationally
inferred best primordial value, $Y_p = 0.232 \pm 0.003 \pm 0.005$ \cite{OSt},
where the errors are 1 $\sigma$ statistical and systematic errors
respectively. Indeed, consistency relies on these errors and even so,
 allows for very little breathing room for any enhancement in
primordial $^4$He.  This is the reason that one can obtain very tight
limits on the number of neutrino flavors \cite{ssg,OSt,kk}.

The $^4$He abundance is primarily determined by the neutron-to-proton
ratio just prior to nucleosynthesis which before the freeze-out of
the weak interaction rates at a temperature $T_f \sim 1$ MeV, is given
approximately by the equilibrium condition
\beq
(n/p) \approx e^{-\Delta m_N/T_f}
\label{eq}
\eeq
where $\Delta m_N = 1.29$ MeV is the neutron-proton mass difference.
(The ratio is slightly altered by free neutron decays between $T_f$
and the onset of nucleosynthesis at about $T \sim 0.1$ MeV.)
Furthermore, freeze-out is determined by the competition
between the weak interaction rates and the expansion rate of the
Universe
\beq
{G_F}^2 {T_f}^5 \sim \Gamma_{\rm wk}(T_f) = H(T_f) \sim \sqrt{G_N N} {T_f}^2
\label{comp}
\eeq
where $N$ counts the total number of relativistic particle
species.
  The presence
of additional neutrino flavors (or any other relativistic species) at
the time of nucleosynthesis increases the overall energy density
of the Universe and hence the expansion rate leading to a larger
value of $T_f$, $(n/p)$, and ultimately $Y_p$.  Because of the
form of eq. (\ref{comp}) it is clear that just as one can place limits
on $N$, any changes in the weak or gravitational coupling constants
can be similarly constrained \cite{yssr}-\cite{acc}.

Constraints on $G_N$ and $G_F$ have often been obtained
under the assumption that these quantities have varied in time as a power-law,
$G \propto t^x$. Constraints on $\delta G/G$ yield an acceptable
range for $x$ \cite{yssr,acc}.  Limits on these couplings as well as
the fine structure constant and neutron-proton mass difference were
considered in \cite{kpw}.  In general, $Y_p$ is most sensitive
to changes in $\Delta m_N$ \cite{kpw}.  It was pointed out
in \cite{ds} however,
that as the Fermi constant can be written directly as the vev of the
Higgs boson in the standard model, $G_F/\sqrt{2} = 1/2v^2$, changes in $G_F$
will naturally induce changes in fermion masses and hence
$\Delta m_N$.  In this context, temporal as well as spatial changes in
$G_F$ were considered in \cite{ss}.

In string theory, the vev of the dilaton field, acts as the string
loop counting parameter \cite{string}. At the (string) tree level,
changes in the vacuum value of the dilaton corresponds directly
to changes in the gravitational coupling $G_N$.  This can
be seen by writing down the action in the string frame \cite{action},
\begin{eqnarray}
S & = & \int d^4x \sqrt{g} e^{-\sqrt{2}\kappa \phi} \left(
{1 \over 2 \kappa^2} R + \partial_\mu \phi \partial^\mu \phi
- {1 \over 2} \partial_\mu y \partial^\mu y -
m_y^2 y^2 \right.\nonumber \\
& & \left. -{\bar \psi} \gamma^\mu D_\mu \psi - m_\psi {\bar \psi} \psi
- \frac{\alpha'}{16 \kappa^2}
F_{\mu\nu}F^{\mu\nu} \right)
\label{sstring}
\end{eqnarray}
where $\phi$ is the dilaton field, $y$ is an arbitrary scalar
field and $\psi$ is an arbitrary fermion.  $D_\mu$ is the
gauge-covariant derivative corresponding to gauge fields with
field strength $F_{\mu\nu}$. $\kappa^2 = 8 \pi G_N$.
However, by performing a conformal transformation, $g_{\mu\nu}
\rightarrow e^{-\sqrt{2} \kappa \phi} g_{\mu\nu}$
we can rewrite (\ref{sstring}) in the Einstein frame as
\begin{eqnarray}
S & = & \int d^4x \sqrt{g}  \left(
{1 \over 2 \kappa^2} R - {1 \over 2} \partial_\mu \phi \partial^\mu \phi
- {1 \over 2} \partial_\mu y \partial^\mu y -
e^{\sqrt{2}\kappa \phi} m_y^2 y^2 \right.\nonumber \\
& & \left. -{\bar \psi} \gamma^\mu D_\mu \psi -
e^{\kappa \phi / \sqrt{2}} m_\psi {\bar \psi} \psi
- \frac{\alpha'}{16 \kappa^2} e^{-\sqrt{2}\kappa\phi}
F_{\mu\nu}F^{\mu\nu}
  \right)
\label{sein}
\end{eqnarray}
Now it apparent that changes in the dilaton vev, will induce
changes in gauge couplings and fermion masses (see details below).
Thus, we will be able to limit changes in the dilaton vev from the
time of nucleosynthesis to today from the observed $^4$He abundance.
As we will see, we will be able to obtain particularly stringent
limits because the dependence on $\phi$ of gauge and Yukawa couplings
induce  changes in quantities such as the Higgs vev, $v$, and $\Lambda_{QCD}$
which are exponentially dependent on the dilaton vev through
the renormalization group equations and dimensional transmutation.
In what follows, we will first derive the general (though approximate)
relations between $Y_p$ and the various couplings and $\Delta m_N$
and the corresponding limits on these quantities.
We will then derive the induced changes in these quantities from changes
in the dilaton vev and hence derive limits on the changes of the
dilaton vev in the context of string theory.  We will also
consider these same effects in the context of Jordan-Brans-Dicke
gravity.

As is well known, the $^4$He abundance is predominantly determined by
the neutron-to-proton ratio and is easily estimated assuming that
all neutrons are incorporated into $^4$He,
\beq
Y_p \approx {2 (n/p) \over 1 + (n/p)}
\eeq
so that
\beq
{\Delta Y \over Y} \approx {1 \over 1 +(n/p)} {\Delta (n/p) \over
(n/p)}
\eeq
{}From eqs.(\ref{eq}) and (\ref{comp}) it is clear that changes in
any of the quantities $G_F, G_N$, or $N$,  will lead to a change
in $T_f$ and hence $Y_p$.
If we keep track of the changes in $T_f$ and $\Delta m_N$
separately, we can write,
\beq
{\Delta (n/p) \over (n/p)} \approx {\Delta m_N \over T_f}
\left( {\Delta T_f \over T_f} - {\Delta^2 m_N \over \Delta m_N} \right)
\eeq
where $\Delta^2 m_N$ is the change in $\Delta m_N$.
Combining these equations we obtain
\beq
{\Delta Y \over Y} \approx
\left( {\Delta T_f \over T_f} - {\Delta^2 m_N \over \Delta m_N} \right)
\label{dy}
\eeq
where the factor $\Delta m_N /T_f (1+(n/p)) \approx 1$.
{}From the consistency of the light elements,
we will take the limit $-0.08 < \Delta Y / Y < 0.01$
assuming a SBBN value of $Y_p = 0.240$ and the observed
range to be $0.221 < Y_p < 0.243$.

As noted above, changes in $T_f$ are induced by changes
in the weak and gravitational couplings and can
be readily determined from (\ref{comp}).  Changes in $\Delta m_N$
can come from a number of sources. One can write the nucleon
mass difference as
\beq
\Delta m_N \sim a \alpha_{em} \Lambda_{QCD} + b v
\label{dm}
\eeq
where a and b are dimensionless constants giving the relative
contributions from the electromagnetic  and weak interactions.
In (\ref{dm}), $v$ is the standard model Higgs expectation value.
A discussion on the contributions to $\Delta m_N$ can be found in
\cite{pn}.  We will take
$a \simeq -.8 {\rm MeV}/{\alpha_{em}}_0 {\Lambda_{QCD}}_0$
where ${\Lambda_{QCD}}_0$ is the present (low energy value)
of $\Lambda_{QCD}$, ${\alpha_{em}^{-1}}_0 \simeq 137$
and $b \simeq 2.1 {\rm MeV} / v_0$ where $v_0$
is the standard value of the Higgs expectation value, $v_0 \simeq 247$
GeV.  Our results will not be particularly sensitive to the
precise values of $a$ and $b$.

In what follows we will consider the effects of changes in gauge
and Yukawa coupling constants. From eq. (\ref{dm}) we see that a
change in the electromagnetic coupling constant will directly
induce a change in $\Delta m_N$.
Changes in the strong coupling constant however can be seen
to have dramatic consequences from the running of the renormalization
group equations\footnote{This observation was made by Dixit and Sher \cite{ds}
in their criticism of the dependence of $Y_p$ on $\alpha_3$
 in \cite{barrow}}.
Indeed the QCD scale $\Lambda$ is determined
by dimensional transmutation
\beq
\alpha_3(M_P^2) \equiv {g_3^2(M_P^2) \over 4 \pi} \approx
{12 \pi \over (33 - 2 N_f)\ln (M_P^2/\Lambda^2)}
\eeq
or for $N_f = 3$
\beq
\Lambda^2 = M_P^2 \exp ({-48\pi^2 \over 27 g_3^2(M_P^2)})
\label{lam}
\eeq
Clearly,  changes in $g_3$ will induce (exponentially) large
changes in $\Lambda_{QCD}$ and therefore in $\Delta m_N$ and $Y_p$.

Similarly, changes in Yukawa couplings can induce large changes in
$\Delta m_N$.  In models in which the electroweak symmetry is broken
radiatively, the weak scale is also determined by dimensional
transmutation \cite{weaktran}. This mechanism is based on the
solution for the renormalization scale at which the Higgs mass$^2$
goes negative, being driven by a Yukawa coupling, presumably
$h_t$.  The weak scale and the Higgs expectation value
then corresponds to the renormalization point and
is given qualitatively by
\beq
v\sim M_P \exp (- 2 \pi c / \alpha_t)
\label{rad}
\eeq
where $c$ is a constant of order 1, and $\alpha_t = h_t^2/4\pi$.
Thus small changes in $h_t$ will induce large changes in $v$
and hence in $T_f$ and $\Delta m_N$.

Let us now look at the implications of a rolling dilaton in string
theory.  We will work in the Einstein frame
and therefore, we will not consider changes in the gravitational coupling
$G_N$. From the form of the action in (\ref{sein}) one can see
that although scalars and fermions have canonical kinetic terms after the
conformal transformation, there remains a dilaton dependence in their masses
as well as in the coefficient of the gauge field strength. From (\ref{sein}),
we define the gauge coupling constant
\beq
{1 \over g^2(M_P^2)} = {\alpha' e^{-\sqrt{2}\kappa\phi} \over 2 \kappa^2}
= {\alpha' S_R \over 2 \kappa}
\eeq
$S$ is the (chiral) multiplet in which the real part
of the scalar is associated with the
dilaton and is used here for convenience. $\alpha'$ is the string tension.
{}From (\ref{lam}), we see that $\Lambda_{QCD}$ is in fact doubly
exponentially dependent on the dilaton\footnote{The dependence
of $\Lambda_{QCD}$ on the dilaton was utilized in a discussion regarding
the dilaton coupling to matter \cite{ekow} and to our
notion of the expansion of the Universe in \cite{CLO}} $\phi$
\beq
\Lambda = M_P \exp \left({-12\pi^2 \alpha' S_R \over 27 \kappa}\right)
\eeq
Therefore we can write the induced change in $\Delta m_N$ as
\beq
{\Delta^2 m_N \over \Delta m_N}
 = {a \alpha \Lambda \over \Delta m_N} \left(
{\Delta \alpha \over \alpha} + {\Delta \Lambda \over \Lambda} \right)
\simeq {-.8 \over 1.3} \left( -{\Delta S_R \over S_R} -
{12 \pi^2 \alpha' \over 27 \kappa} \Delta S_R \right)
\label{dm1}
\eeq

Similarly,  we see from (\ref{sein}) that Yukawa couplings are also
expected to be dilaton dependent.  If we assume that fermion masses
are generated by the Higgs mechanism, then the corresponding Yukawa
term in the Lagrangian would be
\beq
h e^{\kappa \phi / \sqrt{2}} H {\bar \psi} \psi
\label{yuk}
\eeq
and the $\psi$ mass is given by $m_\psi = h e^{\kappa \phi / \sqrt{2}}
\langle
H \rangle = h e^{\kappa \phi / \sqrt{2}} v /\sqrt{2}$.
The effective Yukawa coupling is therefore given by $h e^{\kappa \phi /
\sqrt{2}}$.
Now, as a consequence of eq. (\ref{rad}), the Higgs vev is determined from
\beq
v = M_P \exp \left( {8 \pi^2 c \kappa S_R \over h_t^2} \right)
\eeq
In this case, the induced change in $\Delta m_N$ is
\beq
{\Delta^2 m_N \over \Delta m_N}
 = {b h M_P \over \Delta m_N} {d \over dS_R}\left(
{1 \over \sqrt{S_R}} \exp ({8 \pi^2 c \kappa S_R \over h_t^2}) \right)
\Delta S_R
\simeq - {2.1 \over 1.3} \left({1 \over 2} +
{8 \pi^2 c \kappa S_R \over h_t^2} \right) {\Delta S_R \over S_R}
\label{dm2}
\eeq
Note that in (\ref{dm}) it is really the quark mass difference
which contributes to $\Delta m_N$ so that the dependence is $hv$ rather
than simply $v$.  In addition, changes in $v$ will also induce changes
in the freeze-out temperature $T_f$,
\beq
{\Delta T_f \over T_f} = {4 \over 3} {\Delta v \over v} =
{-32 \pi^2 c \kappa \Delta S_R \over 3 h_t^2}
\label{dt1}
\eeq
Clearly even small changes in the dilaton expectation value
will have dramatic consequences on the $^4$He abundance.

The contributions in eqs. (\ref{dm1},\ref{dm2} and \ref{dt1})
can all be summed to give a net change in $Y_p$. For
$c \sim h_t \sim \kappa S_R \sim 1$ and $g^2 \sim 0.1$, we have from
(\ref{dy})
\beq
{\Delta Y \over Y} \sim 200 \kappa \Delta S_R
\eeq
which in order to be consistent with SBBN gives
\beq
-4 \times 10^{-4}~\la \kappa \Delta S_R~\la 5 \times 10^{-5}
\label{lim}
\eeq
The corresponding limits on other quantities can be easily obtained from
(\ref{lim}) using relations such as $\Delta G_F / G_F \sim
150 \kappa \Delta S_R$, $\Delta \alpha_{em} / \alpha_{em} =
 - \Delta S_R /S_R$, $\Delta h / h =
- \Delta S_R /2 S_R$. This is our main result.

Before concluding, it is interesting to consider these
same limits in the context of Jordan-Brans-Dicke gravity.
If we write down the analogous action to eq. (\ref{sstring})
\begin{eqnarray}
S & = & \int d^4x \sqrt{g}  \left(
 \phi R - {\omega  \over  \phi}
 \partial_\mu \phi \partial^\mu \phi
- {1 \over 2} \partial_\mu y \partial^\mu y -
m_y^2 y^2 \right.\nonumber \\
& & \left. -{\bar \psi} \gamma^\mu D_\mu \psi - m_\psi {\bar \psi} \psi
- \frac{1}{8 g^2}
F_{\mu\nu}F^{\mu\nu} \right)
\label{sjbd}
\end{eqnarray}
and perform the analogous conformal transformation, $g_{\mu\nu}
\rightarrow 2 \kappa^2 \phi g_{\mu\nu}$, we have
\begin{eqnarray}
S & = & \int d^4x \sqrt{g}  \left(
 {1 \over 2 \kappa^2} R -{(\omega + {3 \over 2}) \over 2 \kappa^2 \phi^2}
 \partial_\mu \phi \partial^\mu \phi
- {1 \over 4 \kappa^2 \phi} \partial_\mu y \partial^\mu y -
{1 \over 4 \kappa^4 \phi^2} m_y^2 y^2 \right.\nonumber \\
& & \left. -{1 \over 2 \kappa^2 \phi}
{\bar \psi} \gamma^\mu D_\mu \psi - {1 \over (2 \kappa^2 \phi)^{3/2}}
m_\psi {\bar \psi} \psi
- \frac{1}{8 g^2}
F_{\mu\nu}F^{\mu\nu} \right)
\label{sjbdein}
\end{eqnarray}
Notice first, that the coefficient of $F_{\mu\nu}F^{\mu\nu}$
is independent of $\phi$ and therefore we do not expect that
gauge couplings will vary. This is a result of the conformal
invarience of the gauge kinetic term. Furthermore, notice also that the
fermion and boson kinetic terms are no longer
canonical (as well as that of $\phi$).  If we rescale fermions and bosons
by $(\psi,y) \rightarrow \sqrt{2 \kappa^2 \phi}~(\psi,y)$ and we
assume that masses are generated by the expectation value of a scalar
(through interactions of the form $H^2 y^2$ and $H {\bar \psi} \psi$
and $H$ is similarly rescaled) then we see that the $\phi$ dependence
of the masses drop out.  Therefore, we do not expect effects based
on transdimensional mutation as neither gauge nor Yukawa
couplings will depend on $\phi$ to induce changes in
$\Lambda_{QCD}$ and $v$. The Higgs expectation value will probably
still depend on $\phi$ if its value is determined from a
Higgs potential of the form $V = \lambda H^4 - m^2 H^2$.
If in the standard model $v^2 \sim m^2/\lambda$, then in the conformally
transformed JBD theory, $v^2 \sim m^2 / \lambda (2 \kappa^2 \phi)$.
Thus $G_F \propto v^{-2} \propto \phi$ and $\Delta G_F/ G_F
= \Delta \phi / \phi$.  This could be put in the form of a
constraint on the JBD parameter $\omega$, but will yield
constraints which have been discussed recently in the
literature \cite{JBD} and will not be repeated here.

In summary, we have derived limits on any possible time variation
(from the time of nucleosynthesis to the present) in the
dilaton expectation value (in the context
of string gravity) due to its effect on standard model parameters
such as gauge and Yukawa couplings as well as $\Lambda_{QCD}$ and
the Higgs expectation value from big bang nucleosynthesis.  The induced
variation in the latter two quantities (noting that their
scales are generated through dimensional transmutation)
provides us with stringent limits
on $\Delta S_R$ which can be translated into limits on other couplings.
In the JBD theory of gravity, effects as large were not found as
the JBD scalar does not automatically alter gauge and Yukawa couplings.

\vskip 2.5truecm
\noindent {\bf Acknowledgements}
\vskip 1.0truecm
We would like to thank R. Madden for comments on the manuscript.
This work was supported in part by  the Natural Sciences and
Engineering
Research Council of Canada and by  DOE grant DE-FG02-94ER40823.
%\newpage

\end{document}